\documentstyle[twocolumn,aps,floats]{revtex}

\begin{document}

\draft \tolerance = 10000

\setcounter{topnumber}{1}
\renewcommand{\topfraction}{0.9}
\renewcommand{\textfraction}{0.1}
\renewcommand{\floatpagefraction}{0.9}
\newcommand{\br}{{\bf r}}

\twocolumn[\hsize\textwidth\columnwidth\hsize\csname
@twocolumnfalse\endcsname

\title {Are the Laws of Thermodynamics Consequences of a
Fractal Properties of Universe?}
\author {Kobelev L.Ya.   \\
Department of  Physics, Urals State University \\ Lenina Ave., 51,
Ekaterinburg 620083, Russia  \\ E-mail: leonid.kobelev@usu.ru}

 \maketitle

\begin{abstract}
Why in our Universe the laws of thermodynamics are valid? In the paper is
demonstrated the reason of it: if the time and the space are multifractal
and the Universe is in an equilibrium state the laws of the thermodynamics
are consequences of it's multifractal structure.
\end{abstract}

\pacs{01.30.Tt, 05.45, 64.60.A; 00.89.98.02.90.+p.} \vspace{1cm}

]

\section {Introduction}

It is well known, that the multifractal sets have the characteristics very
similar to the characteristics of a physical quantities (a free energy, an
entropy, a temperature etc)  with which these characteristics can be
formally compared. The connections between the characteristics of
multifractal set and characteristics of physical quantities  formally
correspond to connections between the thermodynamic quantities. This
surprising correspondence till now is completely inexplicable. In the
present paper the multifractal analysis advanced in by Mandelbrot
\cite{1}, \cite{2}, Renyi \cite{3}, Halsy \cite{4}, etc. (see for example,
Rudolf \cite{5}) was used for a substantiation of thermodynamics laws on
the base of the supposition that the space and the time are multifractal
sets. If our Universe state is the state of equilibrium (or the state
nearly equilibrium) the connections between the global characteristics of
Universe as a whole (multifractal time and space) and their local fractal
characteristics will be the same that thermodynamic laws (it is shown on
the basis of the fractal theory of time and space (see \cite{6} -
\cite{8}). From the point of view of the fractal theory of time and space
the thermodynamics relations (as well as thermodynamics in whole) are
consequence of the multifractal structure (structure of time and space) of
our Universe.

    In the theory \cite{6} - \cite{8} the time and the space are
treated as a real physical fields. These fields consist of small
multifractal subsets of time and space ("elements" of time and space), in
turn, approximately treated as "points".  Multifractal sets of time and
space defined on a set of the carrier of a measure $R^{n}$, and contain
all characteristics of the real world by reflected it's in their
fractional dimensions. The fractal dimensions $d_{tr}$ (or $d_{t}$ and
$d_{r}$) in small neighborhood of points $t, r,$ for these "points"
(belonging the sets of $t$) are global dimensions. At the same time for
all space - time continuum these dimensions are local fractal dimensions
(Gelder exponents). The purpose of the paper is the establishment of
connection between the global and the local characteristics of
multifractal space and time on the basis of the multifractal analysis. We
suppose that the state of the Universe (consisting from multifractal time
and spatial sets) may be described as the state close to a thermodynamic
equilibrium. The establishment of such connections enables on to view the
new reason of origin of thermodynamic relations existing in our world,
reducing it to presence of the fractal properties at time and space. We
shall show, that the thermodynamic relations used in physics are a natural
consequence of known mathematical connections between the multifractal
characteristics of the Universe (Universe is considered as multifractal
space - time set described within the framework of the fractal theory of
time and space \cite{6} - \cite{8}). Thus thermodynamics can be considered
as a natural consequence of multifractal characteristics of time and space
of the world in which we live.

\section {Connection Between the Physical and Multifractal Characteristics
in the Multifractal Universe}

Let's consider the Universe as a dynamic system at the  state that close
to a thermodynamic equilibrium (at the present stage of it's development),
defined on a multifractal set $X$. Let the state of the Universe is
characterized by fractal dimensions of a space - time continuum and by
mean values of an internal energy, a free energy, an entropy, a
temperature. If the state of Universe is  close to the thermodynamic
equilibrium, it's characteristics  is possible to describe by a free
energy $F$, entropy $S$, internal energy $E$ and temperature $T$. Let the
Universe has a multifractal nature stipulated by fractional dimensions of
time and space (according to the fractal theory of time and space \cite{6}
- \cite{8}) and is characterized by multifractal set $X(r,t)$  it's space
- time points. The multifractal set $X$ is defined on the carrier of a
measure (set $R^{n}$ with topological dimensions), i.e. $X\ is the subset
of R^{n}$. Let the set of the carrier of a measure is characterized by the
temperature $T_{0}$, by the internal energy $E_{0}=T_{0}$ (in the system
in which the Boltzmann constant is equal to unity), by a free energy
$F_{0}$. Let's define a measure $\mu$ on the set $X$ and consider
connection between invariant scaling characteristics of the multifractal
Universe with a measure $\mu$ on the basis of the theory \cite{6} -
\cite{8} and hypothesis about the origin of the Universe as a result of
explosion (big bang). Because of multifractality of space - time sets, the
scaling transformations (at measuring volume of the Universe with the help
of covering by four- dimension orbs (or cubes) with radius $\delta$), for
example, for mean values of probability of the casual mass distribution
$<p ^{q}>$ (or the random distributions of densities of energy of physical
fields), will look like (see, for example, \cite{4} - \cite{5})

\begin{equation}\label{eq1}
<p^{q}> \sim \delta ^{\tau (q+1)}
\end{equation}

where $q$ is scale factor bound with $q$-dimensions Renyi $dim^{q}_{B}
(X)$ by relation

\begin{equation}\label{eq2}
  dim^{q}_{B}(X)=\frac{\tau (q)}{q-1}
\end{equation}

The dimension Renyi characterizes global scailing characteristics of the
Universe. For definition of it's physical sense we shall consider local
properties of the Universe near to the point $r,t$. The local fractal
dimensions in this point (Gelder's exponent) according to \cite{6} -
\cite{8} looks like

\begin{equation}\label{eq3}
\alpha (x)\equiv d_{t,r}(\vec{r}(t),t)=4+\sum \limits_i \beta _{i}L_{i,t
\vec {r}} (\vec{r} (t),t)
\end{equation}

where $L_{i, t\vec{r}}$  are densities of energy of physical fields in
this point and characterized by the densities of Lagrangians. The quantity
$p_{i}$ in a neighborhood of a point $(r,t)$  is transformed as
\begin{equation}\label{eq4}
  p_{i} \sim \delta ^{d_{tr}(\vec {r}(t),t)}
\end{equation}

From definition of q-dimensions Renyi

\begin{equation}\label{eq5}
 dim^{q}_{B} (\vec {r},t)=\frac{1}{q-1} lim_{\delta \rightarrow 0}
 \frac{log \sum \limits_i p^{q}_{i}}{log \delta}
\end{equation}

follows
\begin{equation}\label{eq6}
  dim^{q}_{B}(\vec {r},t)= \frac {q d(\vec {r}(t),t}{q-1}
\end{equation}

The fractal dimensions $d(r,t)$ in (\ref{eq3}) are the dimensionless
internal energies (after multiplication on $E_{0}$ the relation
(\ref{eq3}) and correspond an internal energies of Universe in a point
with coordinates $(r,t))$ and so, for $q>>1$, follows
\begin{equation}\label{eq7}
 dim^{q}_{B} (\vec {r},t)\approx d(\vec {r} (t),t)
\end{equation}

Therefore the $dim^{q}_{B}(X)$   should has sense of an energies. For
describing of a thermodynamic equilibrium of the Universe there are only
two energies (internal $E$ and free $F$) and $E_{0}$ is bound with $d_{tr}
(r,t)$, therefore the dimensions Renyi there corresponds to a free energy
of the Universe ($F$ divided by $E_{0}$ ) in the $q$ - state. Let's define
now $q$ - state. From (\ref{eq2}) follows, as the Universe cools down  and
also it's temperature is decrease and it's volume grows, that $q$ must
depends on temperature and will increase with Universe cooling. The
simplest dimensionless function satisfying to this requirement is the
function
\begin{equation}\label{eq8}
  T=T_{0}/T
\end{equation}

Now  it is necessary to define a function of state of the Universe - the
entropy $S$. Let's consider subsets $S'(\alpha)$ (of the set $X$) with
identical Gelder's exponents $d_{rt}=\alpha$ (in our case it corresponds
to a selection of an isoenergetic sets of the "internal energy" of the
Universe). In this case joining of subsets $S'(\alpha)$, stratifying
original set $X$, will coincide with the original set. Let's introduce a
spectrum of fractal dimensions $f(\alpha$ ).The joining of all such
subsets makes set $X$. Let's the  fractal dimensions of set $S(\alpha(q))$
(obtained as a result of such stratifying) is $f(\alpha(q))$ (spectrum of
singularities). For each value of $q$ the state of the Universe is
determined as a single-valued state and at alteration $q$ (that is
decreasing  of energy of the Universe because of decreasing of it's
temperature) and expansion of the Universe) function $f(\alpha (q))$  ,
describing scaling properties of set $S(\alpha(q))$, will grows. Such
behavior corresponds to behavior of an entropy (a $q$ - entropy) which we
shall designate by $S$. Hence, to the every state of Universe there
corresponds a spectrum of singularities $f(\alpha(q))$ equal to an $q$ -
entropy $S$.

\section {Connection of a Free Energy and an Entropy as a Consequence of a
Multifractal Nature of the Universe}

We use now the known relation of the multifractal analysis between
q-dimensions Renyi, spectrum of a singularity $f(\alpha(q))$ and local
fractal dimensions $q$, (see, for example, \cite{5})

\begin{equation}\label{eq9}
(q-1) dim^{q}_{B} (X)=q \alpha (q)-f(\alpha (q))
\end{equation}

For $T<<T_{0}$, substituting in (\ref{eq9}) instead of dimensions Renyi
the spectrum of singularities $f(\alpha (q))$, the local fractal
dimensions $\alpha (q)$ and the scaling factor $q$ their physical values
(that we have received earlier) the relation reads

\begin{equation}\label{eq10}
F=E-T S
\end{equation}

The relation (\ref{eq10}) is the basic relation of the thermodynamics. As
relation (\ref{eq10}) is fulfilled for the Universe as a whole, it will be
fulfilled and for it's parts with the state of thermodynamically
equilibrium. Therefore in the Universe with the multifractal time and
space  the realization of the laws of thermodynamics is a simple
consequence of it's structure.

    The analysis of connections of global dimensions and local fractal
characteristics of the fractal space - time carried out above allows to
make the following statements, that are true for a case of equilibrium (or
nearly so by equilibrium) of the state of the Universe:

    a)The free energy of the Universe $F$ can be viewed as fractal $q$ - dimensions
Renyi $(q=T_{0}/T)$ of space - time set $X$ that consist the Universe

\begin{equation}\label{eq11}
dim^{T_{0}/T}_{B} (\vec {r},t)=\frac{T}{T_{0}-T}lim \frac {log (\sum
\limits_i \mu_{i}^{T_{0}/T})}{log \delta}=F
\end{equation}

where $\mu_{i}$ a measure of $i$-th of four-dimensional element of space -
time;

    b) The inverse temperature of the Universe $T_{0}/T$ corresponds to the
$q$-characteristics of the scaling transformation of multifractal  space -
time;

    c) The entropy of Universe $S$ corresponds the spectrum of fractal
dimensions $f(d_{tr}(T_{0}/T))$, defined by dependencies of space-time of
dimensions Renyi $dim^{T_{0}/T}_{B}(\vec {r},t)$ , mean temperature
$T_{0}/T$ and local fractal dimensions of space - time sets with identical
energy $d_{tr}(T_{0}/T)$;

    d) The knowledge of the fractal spectrum and dimensions $d_{tr}(q)$ allows
to find dimensions Renyi from (\ref{eq11}). If the dimensions Renyi is
known, the differentiation (****6.10) with respect to $q$ gives in the
equation

\begin{equation}\label{eq12}
d_{tr}(q)=\frac{d}{dq}[(q-1)dim^{q}_{B}(\vec {r},t)
\end{equation}

It is possible to find, using (\ref{eq9}), the entropy (i,e. the spectrum
of fractal dimensions  $f(T_{0}/T)=f(q))$;

    e) The thermodynamics in viewed model is a consequence of the
multifractality of space - time continuum.

\section{Conclusion}

    The problem of a substantiation of the  thermodynamics within the frame-
work of the fractal theory of time and space presented in this paper, (as
well as a substantiation of irreversibility of time and spatial events
(see \cite{9}) is reduced to a postulating of multifractal properties of
space and time. If model of fractal time and spaces \cite{6} - \cite{8} is
correct, the Universe is open system and  exchange it's energy with the
carrier of a measure $R^{n}$ (or with the alien Universe of inflationary
model \cite{9} or model \cite{10}).

\end{document}